\documentclass[11pt]{llncs}

\usepackage{fullpage}
\usepackage{amsmath}
\usepackage{graphicx}

\usepackage{cite}

\newcommand{\lca}{\ensuremath{\textsc{lca}}}
\newcommand{\rmq}{\ensuremath{\textsc{rmq}}}
\newcommand{\pmrmq}{\ensuremath{\pm 1\rmq}}
\newcommand{\rank}{\ensuremath{\textit{rank}}}
\newcommand{\select}{\ensuremath{\textit{select}}}
\newcommand{\findopen}{\ensuremath{\textit{findopen}}}

\DeclareMathOperator*{\argmin}{argmin}

\title{Optimal Succinctness for Range Minimum Queries}

\author{Johannes Fischer}
\institute{Universit\"at T\"ubingen, Center for Bioinformatics (ZBIT), Sand 14, 72076 T\"ubingen\\
\email{fischer@informatik.uni-tuebingen.de}}

\begin{document}
\maketitle

\begin{abstract}
For a static array $A$ of $n$ totally ordered objects, a \emph{range minimum query} asks for the position of the minimum between two specified array indices. We show how to preprocess $A$ into a scheme of size $2n+o(n)$ bits that allows to answer range minimum queries on $A$ in constant time. This space is asymptotically optimal in the important setting where access to $A$ is not permitted after the preprocessing step. Our scheme can be computed in linear time, using only $n + o(n)$ additional bits for construction. We also improve on LCA-computation in BPS- or DFUDS-encoded trees.
\end{abstract}

\section{Introduction}
For an array $A[1,n]$ of $n$ natural numbers or other objects from a totally 
ordered universe, a \emph{range minimum query} $\rmq_A(i,j)$ for $i\le j$ returns 
the \emph{position} of a minimum element in the sub-array $A[i,j]$;
i.e., $\rmq_A(i,j) = \argmin_{i\le k\le j}\{A[k]\}$. This 
fundamental algorithmic problem has numerous applications, e.g., in text 
indexing \cite{abouelhoda04replacing,sadakane07compressed,fischer08another}, 
text compression \cite{chen07lempel}, document retrieval 
\cite{muthukrishnan02efficient,sadakane07succinct,valimaki07space}, flowgraphs 
\cite{georgiadis04finding}, range queries \cite{saxena09dominance}, position-restricted pattern matching
\cite{crochemore08improved}, just to mention a few.

In all of these applications, the array $A$ in which the range minimum queries (RMQs) are performed 
is static and known in advance, which is also the scenario considered in 
this article. In this case it makes sense to preprocess $A$ into a 
(preprocessing-) \emph{scheme} such that future RMQs can be answered 
quickly. We can hence formulate the following problem.

\begin{problem}[RMQ-Problem]
\begin{description}
\item[Given:] a static array $A[1,n]$ of $n$ totally ordered objects.
\item[Compute:] an (ideally small) data structure, called \emph{scheme}, that allows to answer RMQs on $A$ in constant time.
\end{description}
\end{problem}

The historically first such scheme due to Gabow et
al.\ \cite{gabow84scaling} is based on the following idea: 
because an RMQ-instance can be transformed into an instance of \emph{lowest 
common ancestors} (LCAs) in the \emph{Cartesian Tree} \cite{vuillemin80unifying},
one can use any linear-time preprocessing scheme for 
$O(1)$-LCAs 
\cite{harel84fast,schieber88finding,berkman93recursive,bender05lowest} in 
order to answer RMQs in constant time.


The problem of this transformation \cite{gabow84scaling}, both in theory and in practice, can be 
seen by the following dilemma: storing the Cartesian Tree \emph{explicitly} (i.e., 
with labels and pointers) needs $O(n\log n)$ bits of space, while storing it 
\emph{succinctly} in $2n+o(n)$ bits 
\cite{munro01succinct,benoit05representing} does not allow to map the 
array-indices to the corresponding nodes (see Sect. \ref{sect:type-p} for more details on why this is 
difficult).

A \emph{succinct data structure} uses space that is close to the 
information-theoretic lower bound, in the sense that objects from a universe 
of cardinality $L$ are stored in $(1+o(1)) \log L$ bits.\footnote{Throughout 
this article, space is measured in bits, and $\log$ denotes the binary 
logarithm.} Research on succinct data structures  is very active, and we 
just mention some examples from the realm of trees 
\cite{munro01succinct,benoit05representing,geary06simple,ferragina05structuring,jansson07ultra,sadakane10fully},  
dictionaries \cite{pagh01low,raman07succinct}, and strings 
\cite{grossi05compressed,grossi03high,ferragina05indexing,ferragina07compressed,sadakane03new,sadakane06squeezing},  
being well aware of the fact that this list is far from complete. This 
article presents the first succinct data structure for $O(1)$-RMQs in the 
standard word-RAM model of computation (which is also the model used
in all LCA- and RMQ-schemes cited in this article).

Before detailing our contribution, 
we first classify and summarize existing solutions for $O(1)$-RMQs.

\subsection{Previous Solutions for RMQ}
\begin{table}[t]
\begin{center}
\caption{Preprocessing schemes for $O(1)$-RMQs, where $|A|$ denotes the space for the (read-only) input array.}
\label{tbl:schemes}
\begin{tabular}{|l|c|c|l|}
\hline
reference & $~$final space$~$ & $~$construction space$~$ & comments\\\hline\hline
\cite{harel84fast,schieber88finding,berkman93recursive} & $O(n\log n) + |A|$ & $O(n\log n) + |A|$ & originally devised for LCA, but solve RMQ via Cartesian Tree\\\hline
\cite{bender05lowest} & $O(n\log n) + |A|$ & $O(n\log n) + |A|$ & significantly simpler than 
previous schemes \\\hline
\cite{alstrup02nearest} & $O(n\log n) + |A|$& $O(n\log n) + |A|$ & only solution not 
based on Cartesian Trees \\\hline
\cite{fischer07new} & $2n+o(n) + |A|$ & $2n+o(n) + |A|$ & generalizes to 
$\frac{2}{c}n+o(n) + |A|$ bits, const. $c$ (see Footnote 2)\\\hline
\cite{fischer08practical} & $O(nH_k) + o(n)$ & $2n+o(n) + |A|$ & $H_k$ is the empirical entropy \cite{manzini01analysis} of $A$ (small if $A$ is compressible)\\ \hline\hline
\cite{sadakane07compressed} & $n + o(n)$ & $n + o(n)$ & only for $\pmrmq$; $A$
\emph{must} be encoded as an $n$-bit-vector\\\hline\hline
\cite{sadakane07succinct} & $4n+o(n)$ & $O(n\log n) + |A|$ & only non-systematic data structure so far\\\hline
\textbf{this article} & \boldmath $2n+o(n)$ & \boldmath $3n+o(n) + |A|$ & \textbf{final space requirement optimal}\\\hline
\end{tabular}
\end{center}
\end{table}

In accordance with common nomenclature \cite{gal07cell},
preprocessing schemes for $O(1)$-RMQs can be classified into two different 
types: \emph{systematic} and \emph{non-systematic}. Systematic schemes must store the input array $A$ verbatim
along with the additional information for answering the queries. In such a case the query algorithm
can consult $A$ when answering the queries; this is indeed what all systematic schemes make heavy use of.
On the contrary,
non-systematic schemes must be able to 
obtain their final answer without consulting the array.
This second type is important for at least two 
reasons:
\begin{enumerate}
\item In some applications, e.g., in algorithms for document retrieval 
\cite{muthukrishnan02efficient,sadakane07succinct} or position restricted substring matching \cite{crochemore08improved}, only the \emph{position} 
of the minimum matters, but \emph{not} the value of this minimum. In such 
cases it would be a waste of space (both in theory and in practice) to keep 
the input array in memory, just for obtaining the final answer to the RMQs, 
as in the case of systematic schemes.
\item If the time to access the elements in $A$ is $\omega(1)$, this 
slowed-down access time propagates to the time for answering RMQs if
the query algorithm consults the input array. As a prominent example, in string processing RMQ is often 
used in conjunction with the array of \emph{longest common prefixes} of 
lexicographically consecutive suffixes, the so-called \emph{LCP-array} 
\cite{manber93suffix}. However, storing the LCP-array efficiently in 
$2n+o(n)$ bits \cite{sadakane07compressed} increases the access-time to the 
time needed to retrieve an entry from the corresponding \emph{suffix array} 
\cite{manber93suffix}, which is $\Omega(\log^\epsilon n)$ (constant 
$\epsilon > 0$) at the very best if the suffix array is also stored in 
compressed form \cite{grossi03high,sadakane03new}.
Hence, with a systematic scheme the time 
needed for answering RMQs on LCP could never be $O(1)$ in this 
case. But exactly this would be needed for constant-time navigation in 
RMQ-based compressed suffix trees \cite{fischer08another} (where for 
different reasons the LCP-array is still needed, so this is not the same as 
the above point).
\end{enumerate}
In the following, we briefly sketch previous solutions for RMQ schemes.
For a summary, see Tbl.\ \ref{tbl:schemes}, where, besides 
the final space consumption, in the third column we list the peak space 
consumption at construction time of each scheme, which sometimes differs 
from the former term.

\subsubsection{Systematic Schemes.}
\label{sect:type-a}
Most schemes are based on the Cartesian Tree \cite{vuillemin80unifying}, the 
only exception being the scheme due to Alstrup et al.\ \cite{alstrup02nearest}. 
All direct schemes \cite{bender05lowest,alstrup02nearest,fischer07new,sadakane07compressed}
are based on the idea of splitting the query range into several 
sub-queries, all of which have been precomputed, and then returning the overall
minimum as the final result. The schemes from the first three rows of
Tbl.\ \ref{tbl:schemes} have the same 
theoretical guarantees, with Bender et al.'s scheme \cite{bender05lowest} 
being less complex than the previous ones, and 
Alstrup et al.'s \cite{alstrup02nearest} being even simpler (and most 
practical). The only $O(n)$-bit scheme is due to Fischer and Heun 
\cite{fischer07new} and achieves $2n+o(n)$ bits of space in addition to the space for
the input array $A$. It is based on an ``implicit'' 
enumeration of Cartesian Trees only for very small blocks (instead of the 
whole array $A$). Its further advantage is that it can be adapted to achieve 
entropy-bounds for compressible inputs \cite{fischer08practical}. For 
systematic schemes, no lower bound on space is known.\footnote{The claimed lower 
bound of $2n+o(n)+|A|$ bits under the ``min-probe-model'' \cite{fischer07new} 
turned out to be wrong, as was kindly pointed out to the authors by S.\ Srinivasa 
Rao (personal communication, November 2007). In fact, it is easy to lower 
the space consumption of \cite{fischer07new} to $\frac{2}{c}n+o(n) + |A|$ bits 
(constant integer $c>0$) by grouping $c$ adjacent elements in $A$'s blocks 
together, and ``building'' the Cartesian Trees only on the minima of these 
groups.}

An important special case is Sadakane's $n+o(n)$-bit 
solution \cite{sadakane07compressed} for $\pmrmq$, where it is assumed that 
$A$ has the property that $A[i]-A[i-1]=\pm 1$ for all $1 < i \le n$, and can hence
be encoded as a bit-vector $S[1,n]$, where a `1' at
position $i$ in $S$ indicates that $A$ increases by 1 at position $i$, and
a `0' that it decreases. Because we will make use of this 
scheme in our new algorithm, and also improve on its space consumption in Sect.\ \ref{sect:lowering},
we will describe it in greater detail in Sect.\ \ref{sect:pmrmq}.

\subsubsection{Non-Systematic Schemes.}
\label{sect:type-p}
The only existing scheme is due to Sadakane \cite{sadakane07succinct} 
and uses $4n+o(n)$ bits. It is based on the balanced-parentheses-encoding 
(BPS) \cite{munro01succinct} of the Cartesian Tree $T$ of the input array 
$A$ and a $o(n)$-LCA-computation therein \cite{sadakane07compressed}. The 
difficulty that Sadakane overcomes is that in the ``original'' Cartesian Tree, 
there is no natural mapping between array-indices in $A$ and 
positions of parentheses (basically because 
there is no way to distinguish between left and right nodes in the BPS of 
$T$); therefore, Sadakane introduces $n$ ``fake'' leaves to get such a mapping. 
There are two main drawbacks of this solution.
\begin{enumerate}
\item Due to the introduction of the ``fake'' leaves, it does not achieve the 
\emph{information-theoretic lower bound} (for non-systematic schemes) of $2n-\Theta(\log n)$ bits. This 
lower bound is easy to see because any scheme for RMQs 
allows to reconstruct the Cartesian Tree by iteratively querying the scheme 
for the minimum (in analogy to the definition of the Cartesian Tree); and 
because the Cartesian Tree is binary and each binary tree is a Cartesian 
Tree for some input array, any scheme must use at least 
$\log({2n-1\choose n-1}/(2n-1))=2n-\Theta(\log n)$ bits 
\cite{jacobson89space}.
\item For getting an $O(n)$-time construction algorithm, the (modified)
Cartesian Tree needs to be first constructed in a pointer-based 
implementation, and then converted to the space-saving BPS. This leads to a 
\emph{construction space requirement} of $O(n\log n)$ bits, as 
each node occupies $O(\log n)$ bits in memory. The problem why the BPS 
cannot be constructed directly in $O(n)$ time (at least we are not aware of 
such an algorithm) is that a ``local'' change in $A$ (be it only appending a 
new element at the end) does not necessarily lead to a ``local'' change in the 
tree; this is also the intuitive reason why maintaining dynamic Cartesian Trees is 
difficult \cite{bialynicka06amortized}.
\end{enumerate}

\subsection{Our Results}
\label{sect:results}
We address the two aforementioned problems of Sadakane's solution 
\cite{sadakane07succinct} and resolve them in the following way:
\begin{enumerate}
\item We introduce a new preprocessing scheme for $O(1)$-RMQs that occupies 
only $2n+o(n)$ bits in memory, thus being the first that asymptotically 
achieves the information-theoretic lower bound for non-systematic schemes. The critical reader might 
call this ``lowering the constants'' or ``micro-optimization,'' but we believe that data structures 
using the smallest possible space are of high importance, both in theory and 
in practice. And indeed, there are many examples of this in literature: for 
instance, Munro and Raman \cite{munro01succinct} give a 
$2n+o(n)$-bit-solution for representing ordered trees, while supporting most 
navigational operations in constant time, although a $O(n)$-bit-solution 
(roughly $10n$ bits \cite{munro01succinct}) had already been known for some 10 years before \cite{jacobson89space}. 
Another example comes from compressed text indexing 
\cite{navarro07compressed}, where a lot of effort has been put into 
achieving indexes of size $nH_k + o(n\log \sigma)$ 
\cite{ferragina07compressed}, although indexes of size $O(nH_k) + o(n\log 
\sigma)$ had been known earlier \cite{sadakane03new, 
ferragina05indexing,grossi05compressed}. (Here, $H_k$ is the $k$-th-order 
empirical entropy of the input text $T$ \cite{manzini01analysis} and 
measures the ``compressibility'' of $T$, while $\sigma$ is $T$'s alphabet 
size.)
\item We give a \emph{direct} construction algorithm for the above scheme 
that needs only $n+o(n)$ bits of space in addition to the space for the 
final scheme, thus lowering the construction space for non-systematic schemes from $O(n\log n)$ to 
$O(n)$ bits (on top of $A$). This is a significant improvement, as the space for storing $A$ is not necessarily $\Theta(n\log n)$; for example, if the numbers in $A$ are integers in the range $[1,\log^{O(1)}n]$, $A$ can be stored as an array of packed words using only $O(n\log\log n)$ bits of space. See Sect.\ \ref{sect:app} for a different example. The construction space is an important issue and often 
limits the practicality of a data structure, especially for large inputs (as 
they arise nowadays in web-page-analysis or computational biology).
\end{enumerate}
The intuitive explanation why our scheme works better than Sadakane's 
scheme \cite{sadakane07succinct} is that ours is based on a new tree in which the 
preorder-numbers of the nodes correspond to the array-indices in $A$, 
thereby rendering the introduction of ``fake'' leaves (as described earlier)
unnecessary. In summary, this article is devoted to proving
\begin{theorem}
\label{thm:main}
For an array $A$ of $n$ objects from a totally ordered universe, there is a 
preprocessing scheme for $O(1)$-RMQs on $A$ that occupies only 
$2n+O(\frac{n\log\log n}{\log n})$ bits of memory, while not needing access 
to $A$ after its construction, thus meeting the information-theoretic 
lower bound. This scheme can be constructed in $O(n)$ time, using only $n+o(n)$ bits of 
space in addition to the space for the input and the final scheme.
\end{theorem}

This result is not only appealing in theory, but also important in practice. For example, when RMQs are used in conjunction with sequences of DNA (genomic data), where the alphabet size $\sigma$ is 4, storing the DNA even in \emph{uncompressed} form takes only $2n$ bits, already less than the $4n$ bits of Sadakane's solution \cite{sadakane07succinct}. Hence, halving the space for RMQs leads to a significant reduction of total space. Further, because $n$ is typically very large ($n\approx 2^{32}$ for the human genome), a construction space of $O(n\log n)$ bits is much higher than the $O(n\log\sigma)$ bits for the DNA itself. An additional (practical) advantage of our new scheme is that it also halves the space of the lower order terms (``$o(2n)$ vs.\ $o(4n)$ bits''). This is particularly relevant for realistic problem sizes, where the lower order terms dominate the linear term. An implementation in C++ of our new scheme can be downloaded from \url{http://www-ab.informatik.uni-tuebingen.de/people/fischer/optimalRMQ.tgz}.

\subsection{Outline}
Sect.\ \ref{sect:definitions} presents some basic tools. Sect.\ \ref{sect:new} introduces the new preprocessing scheme. Sect.\ \ref{sect:constr} addresses the linear-time construction of the scheme. Sect.\ \ref{sect:lowering} lowers the second-order term by giving a new data structure for LCA-computation in succinct trees. Sect.\ \ref{sect:app} shows a concrete example of an application where our new preprocessing scheme improves on the total space.

\section{Preliminaries}
\label{sect:definitions}
This section sketches some \emph{known} data structures that we are going to make use of. Throughout this article, we use 
the standard \emph{word-RAM} model of computation, where fundamental arithmetic 
operations on words consisting of $\Theta(\log n)$ consecutive bits can be 
computed in $O(1)$ time.

\subsection{Rank and Select on Binary Strings}
\label{sect:rank}
Consider a \emph{bit-string} $S[1,n]$ of length $n$. We define the 
fundamental \emph{rank}- and \emph{select}-operations on $S$ as follows: 
$\rank_1(S,i)$ gives the number of 1's in the prefix $S[1,i]$, and 
$\select_1(S,i)$ gives the position of the $i$'th 1 in $S$, reading $S$ from 
left to right ($1 \le i \le n$). Operations $\rank_0(S,i)$ and $\select_0(S,i)$ are 
defined similarly for 0-bits. There are data structures of size 
$O(\frac{n\log\log n}{\log n})$ bits in addition to $S$ that support rank- and 
select-operations in $O(1)$ time \cite{munro96tables}.


\subsection{Data Structures for $\pm1$RMQ}
\label{sect:pmrmq}
Consider an array $E[1,n]$ of natural numbers, where the difference between 
consecutive elements in $E$ is either $+1$ or $-1$ (i.e. $E[i]-E[i-1]=\pm 1$ 
for all $1<i\le n)$. Such an array $E$ can be encoded as a bit-vector 
$S[1,n]$, where $S[1]=0$, and for $i>1$, $S[i]=1$ iff $E[i]-E[i-1]=+1$. Then 
$E[i]$ can be obtained by $E[1]+\rank_1(S,i)-\rank_0(S,i)+1 = 
E[1]+i-2\rank_0(S,i)+1$. Under this setting, Sadakane 
\cite{sadakane07compressed} shows how to support RMQs on $E$ in $O(1)$ time, 
using $S$ and additional structures of size $O(\frac{n\log^2\log n}{\log n})$ bits. 
We will improve this space to $O(\frac{n\log\log n}{\log n})$ in Sect.\ \ref{sect:lowering}.
A technical detail is that $\pmrmq(i,j)$ yields the position of the 
\emph{leftmost} minimum in $E[i,j]$ if there are multiple occurrences of 
this minimum.

\subsection{Sequences of Balanced Parentheses}
\label{sect:balanced}
A string $B[1,2n]$ of $n$ opening parentheses `(' and $n$ closing 
parentheses `)' is called \emph{balanced} if in each prefix $B[1,i]$, $1 \le 
i \le 2n$, the number of `)'s is no more than the number of `('s. 
Operation $\findopen(B,i)$ returns the position $j$ of the ``matching'' opening 
parenthesis for the closing parenthesis at position $i$ in $B$. This 
position $j$ is defined as the largest $j<i$ for which 
$\rank_((B,i)-rank_)(B,i) = \rank_((B,j)-rank_)(B,j)$.
The \emph{findopen}-operation can be computed in constant time \cite{munro01succinct}; the most 
space-efficient data structure for this needs $O(\frac{n\log\log 
n}{\log n})$ bits \cite{geary06simple}.

\subsection{Depth-First Unary Degree Encoding of Ordered Trees}
\label{sect:dfuds}
The Depth-First Unary Degree Sequence (DFUDS) $U$ of an ordered tree $T$ is 
defined as follows \cite{benoit05representing}. If $T$ is a 
leaf, $U$ is given by `()'. Otherwise, if the root of $T$ has $w$ subtrees 
$T_1,\dots,T_w$ in this order, $U$ is given by the juxtaposition of $w+1$ 
`('s, a `)', and the DFUDS's of $T_1,\dots,T_w$ in this order, with the 
first `(' of each $T_i$ being omitted. It is easy to see that the 
resulting sequence is balanced, and that it can be interpreted as a 
preorder-listing of $T$'s nodes, where, ignoring the very first `(', a node 
with $w$ children is encoded in \emph{unary} as `$(^w)$' (hence the name DFUDS). 

\section{The New Preprocessing Scheme}
\label{sect:new}
We are now ready to dive into the technical details of our new preprocessing 
scheme. The basis will be a new tree, the \emph{2d-Min-Heap}, defined as 
follows. Recall that $A[1,n]$ is the array to be preprocessed for RMQs. For 
technical reasons, we define $A[0]=-\infty$ as the ``artificial'' overall 
minimum.

\begin{definition}
\label{def:minmin}
The \emph{2d-Min-Heap} $\mathcal{M}_A$ of $A$ is a labeled and ordered tree with 
vertices $v_0, \dots, v_n$, where $v_i$ is labeled with $i$ for all $0 \le i 
\le n$. For $1 \le i \le n$, the parent node of $v_i$ is $v_j$ iff $j < i$, 
$A[j] < A[i]$, and $A[k] \ge A[i]$ for all $j < k \le i$. The order of the 
children is chosen such that their labels are increasing from left to right.
\end{definition}

\begin{figure}[t]
\begin{center}
\includegraphics[scale=1]{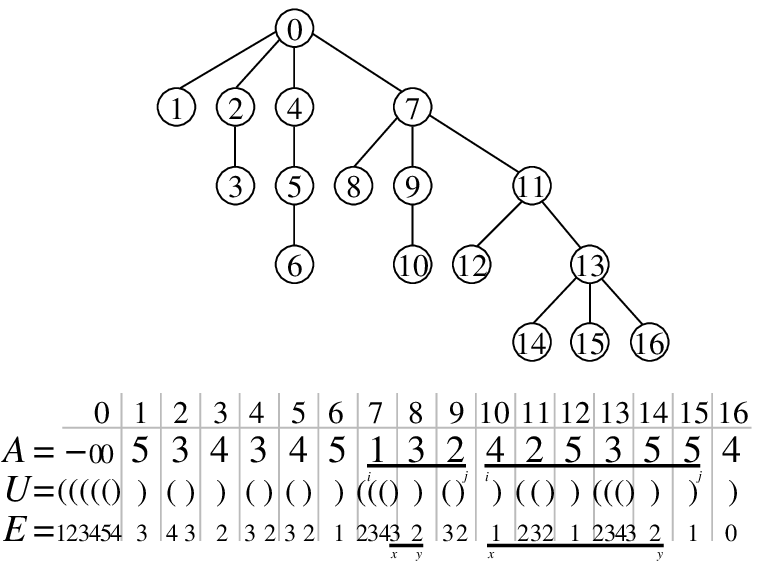}
\end{center}
\caption{Top: The 2d-Min-Heap $\mathcal{M}_A$ of the input array $A$. Bottom: $\mathcal{M}_A$'s DFUDS $U$ and $U$'s excess sequence $E$. Two example queries $\rmq_A(i,j)$ are underlined, including their corresponding queries $\pmrmq_E(x,y)$.}
\label{fig:minmin}
\end{figure}

Observe that this is a well-defined tree with the root being always labeled 
as 0, and that a node $v_i$ can be uniquely identified by its label $i$, 
which we will do henceforth. See Fig.\ \ref{fig:minmin} for an example.

We note the following useful properties of $\mathcal{M}_A$.
\begin{lemma}
\label{lemma:basic}
Let $\mathcal{M}_A$ be the 2d-Min-Heap of $A$.
\begin{enumerate}
\item The node labels correspond to the preorder-numbers of $\mathcal{M}_A$ (counting 
starts at 0).
\item Let $i$ be a node in $\mathcal{M}_A$ with children $x_1,\dots,x_k$. Then $A[i] < 
A[x_j]$ for all $1 \le j \le k$.
\item Again, let $i$ be a node in $\mathcal{M}_A$ with children $x_1,\dots,x_k$. Then 
$A[x_j] \le A[x_{j-1}]$ for all $1 < j \le k$.
\end{enumerate}
\end{lemma}
\textit{Proof.} Because the root of $\mathcal{M}_A$ is always labeled with 0 and the 
order of the children is induced by their labels, property 1 holds. Property 
2 follows immediately from Def.\ \ref{def:minmin}. For property 3, assume 
for the sake of contradiction that $A[x_j] > A[x_{j-1}]$ for two children 
$x_j$ and $x_{j-1}$ of $i$. From property 1, we know that $i < x_{j-1} < 
x_j$, contradicting the definition of the parent-child-relationship in 
$\mathcal{M}_A$, which says that $A[k] \ge A[x_j]$ for all $i < k \le x_j$.\hfill\qed

Properties 2 and 3 of the above lemma explain the choice of the name 
``2d-Min-Heap,'' because $\mathcal{M}_A$ exhibits a minimum-property on both the 
parent-child- and the sibling-sibling-relationship, i.e., in two dimensions.

The following lemma will be central for our scheme, as it gives the desired 
connection of 2d-Min-Heaps and RMQs.
\begin{lemma}
\label{lemma:rmqlca}
Let $\mathcal{M}_A$ be the 2d-Min-Heap of $A$. For arbitrary nodes $i$ and $j$, $1 
\le i < j \le n$, let $\ell$ denote the LCA of $i$ and $j$ in $\mathcal{M}_A$ (recall 
that we identify nodes with their labels). Then if $\ell=i$, $\rmq_A(i,j)$ is 
given by $i$, and otherwise, $\rmq_A(i,j)$ is given by the child of $\ell$ that 
is on the path from $\ell$ to $j$.
\end{lemma}
\textit{Proof.} For an arbitrary node $x$ in $\mathcal{M}_A$, let $T_x$ denote the subtree of 
$\mathcal{M}_A$ that is rooted at $x$. There are two cases to prove.
\begin{description}
\item[$\ell=i.$] This means that $j$ is a descendant of $i$. Due to property 1 
of Lemma \ref{lemma:basic}, this implies that all nodes $i, i+1, \dots, j$ 
are in $T_i$, and the recursive application of property 2 implies that 
$A[i]$ is the minimum in the query range $[i,j]$.
\item[$\ell\ne i.$] Let $x_1,\dots,x_k$ be the children of $\ell$. 
Further, let $\alpha$ and $\beta$ ($1\le\alpha \le \beta \le k$) be defined 
such that $T_{x_\alpha}$ contains $i$, and $T_{x_\beta}$ contains $j$. 
Because $\ell\ne i$ and property 1 of Lemma \ref{lemma:basic}, we must have $\ell 
< i$; in other words, the LCA is not in the query range. But also due to 
property 1, every node in $[i,j]$ is in $T_{x_\gamma}$ for some 
$\alpha\le\gamma\le\beta$, and in particular $x_\gamma\in [i,j]$ for all 
$\alpha < \gamma\le\beta$. Taking this together with property 2, we see that 
$\{x_\gamma : \alpha < \gamma\le\beta\}$ are the only candidate positions 
for the minimum in $A[i,j]$. Due to property 3, we see that $x_\beta$ (the 
child of $\ell$ on the path to $j$) is the position where the overall minimum 
in $A[i,j]$ occurs.\hfill\qed
\end{description}

Note that (unlike for $\pmrmq$) this algorithm yields the \emph{rightmost} 
minimum in the query range if this is not unique. However, it can be easily 
arranged to return the leftmost minimum by adapting the definition of the 
2d-Min-Heap, if this is desired.

To achieve the optimal $2n+o(n)$ bits for our scheme, we represent the 
2d-Min-Heap $\mathcal{M}_A$ by its DFUDS $U$ and $o(n)$ structures for $\rank_)$-, 
$\select_)$-, and $\findopen$-operations on $U$ (see Sect.\ 
\ref{sect:definitions}). We further need structures for $\pmrmq$ on the 
\emph{excess-sequence} $E[1,2n]$ of $U$, defined as $E[i] = \rank_((U,i) - 
\rank_)(U,i)$. This sequence clearly satisfies the property that subsequent 
elements differ by exactly 1, and is already encoded in the right form (by 
means of the DFUDS $U$) for applying the $\pmrmq$-scheme from Sect.\ \ref{sect:pmrmq}.

The reasons for preferring the DFUDS over the BPS-representation
\cite{munro01succinct} of $\mathcal{M}_A$ 
are (1) the operations needed to perform on $\mathcal{M}_A$ are 
particularly easy on DFUDS (see the next corollary), and (2) we have found 
a fast and space-efficient algorithm for constructing the DFUDS directly 
(see the next section).

\begin{corollary}
\label{cor:rmq}
Given the DFUDS $U$ of $\mathcal{M}_A$, $\rmq_A(i,j)$ can be answered in $O(1)$ time 
by the following sequence of operations ($1 \le i < j \le n$).
\begin{enumerate}
\item $x \leftarrow \select_)(U, i+1)$
\item $y \leftarrow \select_)(U,j)$
\item $w \leftarrow \pmrmq_E(x,y)$
\item \emph{if} $\rank_)(U, \findopen(U,w)) = i$ \emph{then return} $i$
\item \emph{else return} $\rank_)(U,w)$
\end{enumerate}
\end{corollary}
\textit{Proof.} Let $\ell$ be the true LCA of $i$ and $j$ in $\mathcal{M}_A$. Inspecting 
the details of how LCA-computation in DFUDS is done \cite[Lemma 
3.2]{jansson07ultra}, we see that after the $\pmrmq$-call in line 3 of the above algorithm, $w+1$ 
contains the starting position in $U$ of the encoding of $\ell$'s child that is on the 
path to $j$.\footnote{In line 1, we correct a minor error in the original article \cite{jansson07ultra} by computing the starting position $x$ 
slightly differently, which is necessary in the case that 
$i=\lca(i,j)$ (confirmed by K.\ Sadakane, personal communication, May 2008).} Line 4 checks if $\ell=i$ by comparing their preorder-numbers 
and returns $i$ in that case (case 1 of Lemma \ref{lemma:rmqlca}) --- it 
follows from the description of the parent-operation in the original article 
on DFUDS \cite{benoit05representing} that this is correct. Finally, in line 
5, the preorder-number of $\ell$'s child that is on the path to $j$ is computed 
correctly (case 2 of Lemma \ref{lemma:rmqlca}).\hfill\qed

We have shown these operations so explicitly in order to emphasize the 
simplicity of our approach. Note in particular that not all operations on 
DFUDS have to be ``implemented'' for our RMQ-scheme, and that we find the correct child
of the LCA $\ell$ directly, without finding $\ell$ explicitly. We encourage the reader 
to work on the examples in Fig.\ \ref{fig:minmin}, where the respective RMQs
in both $A$ and $E$ are underlined and labeled with the variables from
Cor.\ \ref{cor:rmq}.

\section{Construction of 2d-Min-Heaps}
\label{sect:constr}
We now show how to construct the DFUDS $U$ of $\mathcal{M}_A$ in linear time and 
$n+o(n)$ bits of extra space. We first give a general $O(n)$-time algorithm 
that uses $O(n\log n)$ bits (Sect.\ \ref{sect:constr1}), and then show how 
to reduce its space to $n+o(n)$ bits, while still having linear running time 
(Sect.\ \ref{sect:constr2}).

\subsection{The General Linear-Time Algorithm}
\label{sect:constr1}
We show how to construct $U$ (the DFUDS of $\mathcal{M}_A$) in linear time. The idea 
is to scan $A$ from \emph{right to left} and build $U$ from right to left, 
too. Suppose we are currently in step $i$ ($n \ge i \ge 0$), and $A[i+1,n]$ 
have already been scanned. We keep a stack $S[1,h]$ (where $S[h]$ is the 
top) with the properties that $A[S[h]]\ge \dots \ge A[S[1]]$, and $i < S[h] 
< \dots < S[1] \le n$.
$S$ contains exactly those indices $j\in[i+1,n]$ for which $A[k]\ge A[j]$
for all $i < k < j$.
Initially, both $S$ and $U$ are empty. When in step $i$, we first write a 
`)' to the current beginning of $U$, and then pop all $w$ indices from $S$ 
for which the corresponding entry in $A$ is strictly greater than $A[i]$. To 
reflect this change in $U$, we write $w$ opening parentheses `(' to the 
current beginning of $U$. Finally, we push $i$ on $S$ and move to the next 
(i.e. preceding) position $i-1$. It is easy to see that these changes on $S$ 
maintain the properties of the stack.
If $i=0$, we write an initial `(' to $U$ and stop the algorithm.

The correctness of this algorithm follows from the fact that due to the 
definition of $\mathcal{M}_A$, the degree of node $i$ is given by the number $w$ of 
array-indices to the right of $i$ which have $A[i]$ as their closest smaller 
value (properties 2 and 3 of Lemma \ref{lemma:basic}). Thus, in $U$ node $i$ is encoded as `$(^w)$', which is exactly what 
we do. Because each index is pushed and popped exactly once on/from $S$, the 
linear running time follows.

\subsection{$O(n)$-bit Solution}
\label{sect:constr2}
The only drawback of the above algorithm is that stack $S$ requires $O(n\log 
n)$ bits in the worst case. We solve this problem by representing $S$ as a 
\emph{bit-vector} $S'[1,n]$. $S'[i]$ is 1 if $i$ is on $S$, and 0 otherwise. 
In order to maintain constant time access to $S$, we use a standard 
blocking-technique as follows. We logically group $s=\lceil\frac{\log n}{2}\rceil$ 
consecutive elements of $S'$ into \emph{blocks} $B_0,\dots,B_{\lfloor\frac{n-1}{s}\rfloor}$.
Further, $s'=s^2$ elements are grouped into
\emph{super-blocks} $B'_0,\dots,B'_{\lfloor\frac{n-1}{s'}\rfloor}$.

For each such (super-)block $B$ that contains at least one 
1, in a new table $M$ (or $M'$, respectively) at position $x$ we store the block number of the 
leftmost (super-)block to the right of $B$ that contains a 1, in $M$ only relative to
the beginning of the super-block. These tables 
need $O(\frac{n}{s}\log (s'/s)) = O(\frac{n\log\log n}{\log n})$ 
and $O(\frac{n}{s'}\log (n/s)) = O(\frac{n}{\log n})$ 
bits of space, respectively. Further, for all possible bit-vectors of length $s$ we 
maintain a table $P$ that stores the position of the leftmost 1 in that 
vector. This table needs $O(2^s \cdot \log s) = O(\sqrt{n}\log\log n)=o(n)$ 
bits. Next, we show how to use these tables for constant-time access to $S$, 
and how to keep $M$ and $M'$ up to date.

When entering step $i$ of the algorithm, we known that $S'[i+1]=1$, because 
position $i+1$ has been pushed on $S$ as the last operation of the previous 
step. Thus, the top of $S$ is given by $i+1$. For finding the leftmost 1 in 
$S'$ to the right of $j>i$ (position $j$ has just been popped from $S$), we 
first check if $j$'s block $B_x$, $x=\lfloor\frac{j-1}{s}\rfloor$, contains 
a 1, and if so, find this leftmost 1 by consulting $P$. If $B_x$ does not 
contain a 1, we jump to the next block $B_{y}$ containing a 1 by 
first jumping to $y=x+M[x]$, and if this block does not contain a 1, by further jumping to 
$y=M'[\lfloor\frac{j-1}{s'}\rfloor]$. In block $y$, we can again use $P$ to find the leftmost 1. Thus, we can find 
the new top of $S$ in constant time.

In order to keep $M$ up to date, we need to handle the operations where (1) 
elements are pushed on $S$ (i.e., a 0 is changed to a 1 in $S'$), and (2) 
elements are popped from $S$ (a 1 changed to a 0). Because in step $i$
only $i$ is pushed on $S$, for operation (1) we just need to 
store the block number $y$ of the former top in $M[x]$ 
($x=\lfloor\frac{i-1}{s}\rfloor$), if this is in a different block (i.e., if 
$x \ne y$). Changes to $M'$ are similar. For operation (2), nothing has to be done at all, because even 
if the popped index was the last 1 in its (super-)block, we know that 
all (super-)blocks to the left of it do not contain a 1, so no values in $M$ and $M'$ have to
be changed. Note that this only works because elements to the right of $i$
will never be pushed again onto $S$. This completes the description of the
$n+o(n)$-bit construction algorithm.

\section{Lowering the Second-Order-Term}
\label{sect:lowering}
Until now, the second-order-term is dominated by the $O(\frac{n\log^2\log 
n}{\log n})$ bits from Sadakane's preprocessing scheme for $\pmrmq$ (Sect.\ 
\ref{sect:pmrmq}), while all other terms (for $\rank, \select$ and 
$\findopen$) are $O(\frac{n\log\log n}{\log n})$. We show in this section a 
simple way to lower the space for $\pmrmq$ to $O(\frac{n\log\log n}{\log 
n})$, thereby completing the proof of Thm.\ \ref{thm:main}.

As in the original algorithm \cite{sadakane07compressed}, we divide the input array $E$ into 
$n'=\lfloor\frac{n-1}{s}\rfloor$ blocks of size $s=\lceil\frac{\log n}{2}\rceil$.
Queries are decomposed into at most three non-overlapping sub-queries, where the first 
and the last sub-queries are inside of the blocks of size $s$, and the 
middle one exactly spans over blocks. The two queries inside of the blocks 
are answered by table lookups using $O(\sqrt{n}\log^2n)$ bits, as in the 
original algorithm.

For the queries spanning exactly over blocks of size $s$, we proceed as 
follows. Define a new array $E'[0,n']$ such 
that $E'[i]$ holds the minimum of $E$'s $i$'th block.
$E'$ is represented only \emph{implicitly} by an array $E''[0,n']$, where
$E''[i]$ holds the position of the minimum in the $i$'th block, relative to the
beginning of that block. Then $E'[i] = E[is+E''[i]]$.
Because $E''$ stores $n/\log n$ numbers from the range $[1,s]$,
the size for storing $E'$ is thus $O(\frac{n\log\log n}{\log n})$ bits.
Note that unlike $E$, $E'$ does not necessarily 
fulfill the $\pm 1$-property.
$E'$ is now preprocessed for constant-time RMQs with the systematic scheme of 
Fischer and Heun \cite{fischer07new}, using $2n'+o(n') = 
O(\frac{n}{\log n})$ bits of space. Thus, by querying $\rmq_{E'}(i,j)$ for $1\le i\le j \le n'$, we can also find the minima for 
the sub-queries spanning exactly over the blocks in $E$.

Two comments are in 
order at this place. First, the used RMQ-scheme \cite{fischer07new} does allow the 
input array to be represented implicitly, as in our case. And second, it
does not use Sadakane's solution for $\pmrmq$, so 
there are no circular dependencies.

As a corollary, this approach also lowers the space for LCA-computation in
BPS \cite{sadakane07compressed} and DFUDS 
\cite{jansson07ultra} from $O(\frac{n\log^2\log n}{\log n})$ to 
$O(\frac{n\log\log n}{\log n})$, as these are based on $\pmrmq$:
\begin{corollary}
\label{thm:lca}
Given the BPS or DFUDS of an ordered tree $T$, there is a data structure of size $O(\frac{n\log\log n}{\log n})$ bits that allows to answer LCA-queries in $T$ in constant time.
\end{corollary}

\section{Application in Document Retrieval Systems}
\label{sect:app}
We now sketch a concrete example of where Thm.\ \ref{thm:main} lowers the construction space of a different data structure. This section is meant to show that there are indeed applications where the memory bottleneck is the construction space for RMQs. We consider the following problem:

\begin{problem}[Document Listing Problem \cite{muthukrishnan02efficient}]
\begin{description}
\item[Given:] a collection of $k$ text documents $\mathcal{D}=\{D_1,\dots, D_k\}$ of total length $n$.
\item[Compute:] an index that, given a search pattern $P$ of length $m$, returns all $d$ documents from $\mathcal{D}$ that contain $P$, in time proportional to $m$ and $d$ (in contrast to \emph{all} occurrences of $P$ in $\mathcal{D}$).
\end{description}
\end{problem}

Sadakane \cite[Sect.\ 4]{sadakane07succinct} gives a succinct index for this problem. It uses three parts, for convenience listed here together with their \emph{final} size:
\begin{itemize}
\item Compressed suffix array \cite{sadakane03new} $A$ of the concatenation of all $k$ documents, $|A|=\frac{1}{\epsilon}H_0n + O(n)$ bits.
\item Array of document identifiers $D$, defined by $D[i]=j$ iff the $A[i]$'th suffix ``belongs to'' document $j$. Its size is $O(k\log \frac{n}{k})$ bits
\item Range minimum queries on an array $C$, $|RMQ|=4n+o(n)$ bits. Here, $C$ stores positions in $A$ of nearest previous occurrences of indexed positions from the \emph{same} document, $C[i] = \max\{j<i : D[j]=D[i]\}$. In the query algorithm, only the \emph{positions} of the minima matter; hence, this is a \emph{non-systematic} setting.
\end{itemize}

Apart from halving the space for RMQ from $4n$ to $2n$ bits, our new scheme also lowers the peak space consumption of Sadakane's index for the Document Listing Problem. Let us consider the \emph{construction} time and space for each part in turn:

\begin{itemize}
\item Array $A$ can be built in $O(n)$ time and $O(n)$ bits (constant alphabet), or $O(n\log\log|\Sigma|)$ time using $O(n\log|\Sigma|)$ bits (arbitrary alphabet $\Sigma$) of space \cite{hon09breaking}.
\item Array $D$ is actually implemented as a fully indexable dictionary \cite{raman07succinct} called $D'$, and can certainly be built in linear time using $O(n)$ bits working space, as we can always couple the block-encodings \cite{raman07succinct} with the $o(n)$-bit structures for uncompressed solutions for rank and select \cite{munro96tables}.
\item As already mentioned before, for a fast construction of Sadakane's scheme for $O(1)$-RMQs on $C$, we would have needed $\Theta(n\log n)$ bits. Our new method lowers this to $O(n)$ bits construction space. Note that array $C$ needs never be stored \emph{plainly} during the construction: because $C$ is scanned only once when building the DFUDS (Sect.\ \ref{sect:constr}) and is thus accessed only sequentially, we only need to store the positions in $A$ of the last seen document identifier for each of the $k$ documents. This can be done using a plain array, so $|C|=O(k\log n)$ bits.
\end{itemize}

In summary, we get:

\begin{theorem}
\label{thm:retrieval}
The construction space for Sadakane's Index for Document Listing \cite{sadakane07succinct} is lowered from $O(n\log n)$ bits to $O(n + k\log n)$ bits (constant alphabet) or $O(n\log|\Sigma| + k\log n)$ bits (arbitrary alphabet $\Sigma$) with our scheme for RMQs from Thm.\ \ref{thm:main}, while not increasing the construction time.
\end{theorem}

This is especially interesting if $k$, the number of documents, is not too large, $k=O(\frac{n \log|\Sigma|}{\log n})$.

\section{Concluding Remarks}
We have given the first optimal preprocessing scheme for $O(1)$-RMQs under 
the important assumption that the input array is not available after 
preprocessing. To the expert, it might come as a surprise that our 
algorithm is \emph{not} based on the Cartesian Tree, a concept that has 
proved to be very successful in former schemes. Instead, we have introduced 
a new tree, the 2d-Min-Heap, which seems to be better suited for our task.\footnote{The Cartesian Tree and the 2d-Min-Heap are certainly related, as they are both obtained from the array, and it would certainly be possible to derive the 2d-Min-Heap (or a related structure obtained from the natural bijection between binary and ordered rooted trees) from the Cartesian Tree, and then convert it to the BPS/DFUDS. But see the second point in Sect.\ \ref{sect:results} why this is \emph{not a good idea.}}
We hope to have thereby introduced a new versatile data structure to the algorithms community. And indeed, we are already aware of the fact that the 2d-Min-Heap, made public via a preprint of this article \cite{fischer08optimal}, is pivotal to a new data structure for succinct trees \cite{sadakane10fully}.

We leave it as an open research problem whether the $3n+o(n)$-bit construction space be lowered to an optimal 
$2n+o(n)$-bit ``in-place'' construction algorithm. (A simple example shows that it is \emph{not} possible to use the leading $n$ bits of the DFUDS for the stack.)

\section{Recent Developments}
It has recently been shown \cite{gogXXadvantages} that replacing the $\pmrmq$-call in Cor.\ \ref{cor:rmq} by the \emph{range restricted enclose}-operation is advantegeous in practice, as this latter operation can be implemented by \emph{sharing} the most consuming parts of the data structures with those of the findopen-operation.

\subsection*{Acknowledgments}
The author wishes to thank Volker Heun, Veli M\"akinen, Gonzalo Navarro, and the anonymous LATIN'10-referees for their helpful comments on this article.

\bibliographystyle{abbrv}
\bibliography{literatur}

\end{document}